# Multidimensional On-lattice Higher-order Models in the Thermal Lattice Boltzmann Theory

Jae Wan Shim

*KIST and University of Science and Technology, 136-791, Seoul, Korea*

**Abstract.** We present a set of uniform polynomial equations that provides multidimensional *on-lattice* higher-order models of the lattice Boltzmann theory, while keeping compact the number of discrete velocities. As examples, we explicitly derive two- and three-dimensional on-lattice models applicable to describing thermal compressible flows of the accuracy levels of the Navier-Stokes equations with smaller numbers of discrete velocities in comparison to the existing models. We demonstrate the accuracy and stability of the three-dimensional model by using the Riemann problem.

Since the introduction of Ulam and von Neumann's concept of cellular automata, applications have been developed in diverse areas such as biology, chemistry, physics, and astronomy [1, 2]. In the field of fluid flows, Frisch, Hasslacher, and Pomeau introduced lattice-gas automata [3-5] as a model of the Navier-Stokes equations, which led to the development of the lattice Boltzmann theory [6-10]. Various attempts have been recently made to derive higher-order accuracy models classified as *on-lattice* (space-filling) or off-lattice (non-space-filling) types according to the need of the additional interpolation caused by the mismatch between the positions of the lattice nodes and fluid particles. Conventional methods based on the Gauss-type quadrature and the Hermite polynomial expansion, including [11], give either on-lattice lower-order or off-lattice higher-order models. Therefore, a framework providing multidimensional on-lattice higher-order models has been demanded for the efficiency of higher-order models. Prior to looking at the previous works to obtain higher-order models, let us first define the accuracy order of a model by its velocity momenta, which give macroscopic physical properties such as density, flow velocity, temperature, and heat flux *etc.*, so that, if a model has $m$th-order accuracy, it gives the velocity momenta of the Maxwellian up to the $m$th-order. Yudistiawan *et al*. provided off-lattice models, including a 3D 27-velocity model, which needs the additional interpolation work and only has third-order accuracy or the level of the isothermal Navier-Stokes equations [12]. Chikatamarla and Karlin [13] developed a relation between entropy construction and roots of Hermite polynomials, and gave on-lattice, 1D five-velocity models, which can be obtained by other ways, including [14], also having third-order accuracy. The same authors extended their work to 3D space and obtained a 41-velocity model with equal accuracy [15]. They also gave 1D seven- and eleven-velocity models in the appendix of [15]; however, they provided neither corresponding equilibrium distributions to increase actually the order of accuracy nor pruned models to reduce the number of discrete velocities while keeping their order of accuracy. Philippi *et al.* obtained on-lattice 2D models, including a 37-velocity model satisfying the fourth-order accuracy, the level of thermal compressible Navier-Stokes equations [16]. Their method is based on finding the inner product that preserves the norm and the orthogonality of the Hermite polynomial tensors in Hilbert space. Although they provide the on-lattice fourth-order accuracy model, their work is limited to 2D space, and their comment that the 37-velocity model is the minimal square lattice giving that accuracy is not correct; as a counterexample, we will present a 33-velocity model with equal accuracy. Surmas *et al*. [17] obtained 3D models based on the work of [16], including the 3D 107-velocity model of fourth-order accuracy. Nie *et al.* and Shan respectively presented the 3D 121- and 103-velocity models of fourth-order accuracy by using a relation obtained from Gauss-Hermite quadrature which has off-lattice nature [18, 19]. However, we will present an analytical approach having on-lattice nature and give a 95-velocity model with equal accuracy, which is fewer in the number of discrete velocities. Rubinstein and Luo applied the group theory to obtaining on-lattice higher-order models but they provided models having accuracy up to the third-order level [20].

Here, we present a set of polynomial equations in a single form, which is an extension of the univariate polynomial equation providing 1D higher-order on-lattice models in [21], with a way of counting the number of equations for a given order of accuracy by the partition in number theory. This enables us to obtain systematically multidimensional on-lattice higher-order models for the purpose of increasing efficiency by reducing the number of discrete velocities while keeping accuracy. We explicitly derive 2D and 3D models having smaller numbers of discrete velocities than the previously known results and show the stability and accuracy of the 3D model by using the Riemann problem.

Fluid flows described by the lattice Boltzmann equation use the notion of fictitious particles moving their positions and changing their velocity distribution according to a simple rule $f_i(\mathbf{x}+\mathbf{V}_i\Delta t,t+\Delta t)=(1-\omega)f_i(\mathbf{x},t)+\omega f_i^{eq}(\mathbf{x},t)$ where $f_i(\mathbf{x},t)$ is the density distribution of particles having discrete velocities $\mathbf{V}_i$ at position $\mathbf{x}$ at time $t$. The $f_i^{eq}(\mathbf{x},t)$ is the equilibrium state of $f_i(\mathbf{x},t)$ and $\omega$ adjusts viscosity. For the continuous velocity space, whereas our goal is to obtain $f_i^{eq}(\mathbf{x},t)$ in a discrete velocity space, the equilibrium state is the Maxwellian

$$f^{eq}(\mathbf{V};\rho,\mathbf{U},T) = \rho \left(\frac{m_g}{2\pi k_B T}\right)^{D/2} \exp\left(-\frac{m_g \|\mathbf{V}-\mathbf{U}\|^2}{2k_B T}\right)$$

where $\mathbf{V}$ is the particle velocity, $\rho \equiv \rho(\mathbf{x},t)$ the density, $\mathbf{U} \equiv \mathbf{U}(\mathbf{x},t)$ the flow velocity, $T \equiv T(\mathbf{x},t)$ the temperature, $D$ the dimension of space, $k_B$ the Boltzmann constant, and $m_g$ the molecular mass. Dimensionless variables are defined by $\theta \equiv T/T_0$, $\mathbf{v} \equiv \Theta^{-1/2}\mathbf{V}$, and $\mathbf{u} \equiv \Theta^{-1/2}\mathbf{U}$ where $\Theta \equiv 2k_B T_0/m_g$ for convenience.

We use the following constraints of velocity momenta to obtain $f_i^{eq}(\mathbf{x},t)$;

$$\mathbf{M}(n) \equiv \int \mathbf{v}^n f^{eq}(\mathbf{v})d\mathbf{V} = \sum_i \mathbf{v}_i^n f_i^{eq}(\mathbf{v}_i) \text{ for } n=0,1,\ldots,m \qquad (1)$$

where $\mathbf{v}^n \equiv v_{x_1}^{a_1} v_{x_2}^{a_2} \ldots v_{x_D}^{a_D}$ and $\mathbf{v}_i^n \equiv v_{i,x_1}^{a_1} v_{i,x_2}^{a_2} \ldots v_{i,x_D}^{a_D}$ such that $a_1+a_2+\ldots+a_D=n$ and $a_\alpha \in \mathbb{N}_0$ for $\alpha=1,2,\ldots,D$. The subscripts $x_i$ for $i=1,2,\ldots,D$ signify the coordinates in $D$-dimensional Cartesian coordinate system. In Formula (1), $m$ is a factor determining the accuracy of a model. The physical meaning of Formula (1) is the conservation of physical properties such as mass, momentum, pressure tensor, energy flux, and the change rate of the energy flux *etc.* as the order $n$ of **M** increases. We can express $f^{eq}$ by series expansions $f_E^{eq}$ such as the Hermite and the Taylor having the form of $f_E^{eq}(\mathbf{v};\rho,\mathbf{U},T) = \exp(-v^2)P^{(N)}(\mathbf{v})$ where $P^{(N)}(\mathbf{v})$ is a polynomial of degree $N$ in $\mathbf{v}$ and $v^2 \equiv \mathbf{v}\cdot\mathbf{v}$ [22]. Note that $\rho$, **U**, and $T$ are imbedded in the coefficients of $P^{(N)}(\mathbf{v})$. We do not confine the expansions to the Hermite because the Taylor expansion gives various results including the Hermite, according to the choice of infinitesimals [22]. When a Taylor expansion is used, advantage is observed in stability [21]. The order of series expansions of $f^{eq}$ is another factor determining the accuracy of a model along with the maximum order $m$ of **M**. By analogy with $f_E^{eq}$, we expect $f_i^{eq}$ in the form of $f_i^{eq}(\mathbf{v}_i) = w_i\, P^{(N)}(\mathbf{v}_i)$ where $w_i$ is a constant weight coefficient into which $\exp(-v_i^2)$ is merged. With the use of polynomial representations in $f_E^{eq}$ and $f_i^{eq}$, Formula (1) can be written by

$$\int \exp(-v^2) P^{(k+N)}(\mathbf{v})\, d\mathbf{v} = \sum_i w_i P^{(k+N)}(\mathbf{v}_i) \text{ for } k = 0,1,...,m. \qquad (2)$$

or

$$\sum_{n=0}^{k+N} c_n \left\{ \int \exp(-v^2)\mathbf{v}^n\, d\mathbf{v} - \sum_i w_i \mathbf{v}_i^n \right\} = 0 \text{ for } k = 0,1,...,m$$

by the definition of $P^{(k+N)}(\mathbf{v}) = \sum_{n=0}^{k+N} c_n \mathbf{v}^n$. Therefore, for any coefficient $c_n$, Formula (2) is satisfied if and only if

$$\int \exp(-v^2)\mathbf{v}^n d\mathbf{v} = \sum_i w_i \mathbf{v}_i^n \text{ for } n = 0,1,...,m+N. \qquad (3)$$

The left side of Formula (3) can be evaluated by

$$\int \exp(-v^2)\mathbf{v}^n\, d\mathbf{v} = \begin{cases} \prod_{\alpha=1}^{D} \Gamma((a_\alpha+1)/2) & \text{when even for all } a_\alpha, \\ 0 & \text{otherwise} \end{cases} \qquad (4)$$

where $\Gamma$ is the Gaussian Gamma function expressed by the double factorial as $\Gamma((n+1)/2) = \sqrt{\pi}(n-1)!!/2^{n/2}$. On the analogy of the Maxwellian, $f_i^{eq}$ is isotropic when $\mathbf{u} = \mathbf{0}$, so is the set of $\mathbf{v}_i$. Then Formula (3) is satisfied when $n$ is odd and any $a_\alpha$ is odd although $n$ is even, hence it can be written by

$$\sum_i w_i \mathbf{v}_i^n = \prod_{\alpha=1}^{D} \Gamma((a_\alpha+1)/2) \text{ for } n=0,2,4,\ldots,n_{\max} \text{ and even } a_\alpha \qquad (5)$$

where $n_{\max}$ is equal to $m+N$ when it is even or to $m+N-1$ when odd. Note that $m$ and $N$ are two factors determining the accuracy of the discrete models. If we expand the Maxwellian by the Hermite expansion $f_{HE}^{eq}$ and want to satisfy **M** up to the $m$ th-order in Formula (1), it is required $N \geq m$ [21,22].

To count the number of equations provided by Formula (5), we introduce the partition in number theory. A partition $\pi(q)$ is a number of ways to represent a positive integer $q$ as a sum of natural numbers with $\pi(0)=1$. Its generating function is written by [23]

$$\sum_{q=0}^{\infty} \pi(q) x^q = \prod_{k=1}^{\infty} Q(k) \equiv \prod_{k=1}^{\infty} \frac{1}{1-x^k} \text{ for } |x|<1. \qquad (6)$$

To prove Formula (6), let us express $\prod_{k=1}^{\infty} Q(k)$ by using geometric series expansions, $Q(k) = 1 + x^k + x^{2k} + x^{3k} + \cdots$. Then, the ways of obtaining $x^q$ by multiplying each terms, $x$ to the power of $c_k k$, extracted from each $Q(k)$ is the possible combinations of $c_k$ in $q = \sum_{k=1}^{\infty} c_k k$, which is equivalent to $\pi(q)$. A partition $\pi(q,D)$ restricted by the number of parts $D$ is equivalent to a partition restricted by the largest part $D$. This is proven by interchanging the rows and columns of the Ferrers graph [23]. By using this and the proof of Formula (6), it is easy to obtain

$$\sum_{q=0}^{\infty} \pi(q,D) x^q = \prod_{k=1}^{D} \frac{1}{1-x^k} \text{ for } |x|<1. \qquad (7)$$

By $\bar{\pi}(2q,D)$, we define partitions restricted by the number of parts $D$ and the parts themselves are even. Then we have $\pi(q,D) = \bar{\pi}(2q,D)$ because we obtain $\bar{\pi}(2q,D)$ by multiplying two to all the parts of $\pi(q,D)$. Consequently, the number of equations provided by Formula (5) up to $n_{\max}=2m$ can be calculated by

$$\Omega(m,D) \equiv \sum_{q=0}^{m} \bar{\pi}(2q,D) = \sum_{q=0}^{m} \pi(q,D). \qquad (8)$$

A solution satisfying Formula (5) up to $n_{max} = 2m$ is a model of $m$ th-order accuracy if $m$ th-order $f_{HE}^{eq}$ is used [21,22]. For convenience, we give Table 1 for the values of $\Omega(m,D)$. For example, a thermal compressible flow of the level of the Navier-Stokes equations needs the accuracy of $m = 4$ when we use $m$ th-order $f_{HE}^{eq}$. In this case, the numbers of equations for two- and three-dimensional spaces are 9 and 11, respectively, as in Table 1.

|     | $m=0$ | $m=1$ | $m=2$ | $m=3$ | $m=4$ | $m=5$ | $m=6$ | $m=7$ |
|-----|-------|-------|-------|-------|-------|-------|-------|-------|
| $D=1$ | 1 | 2 | 3 | 4 | 5 | 6 | 7 | 8 |
| $D=2$ | 1 | 2 | 4 | 6 | 9 | 12 | 16 | 20 |
| $D=3$ | 1 | 2 | 4 | 7 | 11 | 16 | 23 | 31 |
| $D=4$ | 1 | 2 | 4 | 7 | 12 | 18 | 27 | 38 |

**Table 1**. The number of equations $\Omega(m,D)$ with respect to the order of accuracy $m$ up to eight and the space dimension $D$ up to four when we use $m$ th-order $f_{HE}^{eq}$ for the equilibrium distribution.

To count the number of discrete velocities, which satisfies the $m$th-order accuracy, let us consider, firstly 2D space, then 3D space. To satisfy the isotropic condition, we generate discrete velocities from a representative one by symmetry. Then we have a group of eight elements $G_8$ generated by $c(i_\alpha, j_\alpha)$ with a non-zero constant $c$ and two different natural numbers $i_\alpha$ and $j_\alpha$, a group of four elements $G_4$ by $c(i_\alpha, 0)$ or $c(i_\alpha, i_\alpha)$, or a group of one element $G_1$ generated by (0, 0). Each group shares weight coefficient $w_i$ for isotropy. Accordingly, we have the total number of discrete velocities $n(\mathbf{v}_i) = n_1 + 4n_4 + 8n_8$ where $n_i$ is the number of $G_i$ and the number of unknown variables is $n_1 + n_4 + n_8 + 1$ for a given set of $i_\alpha$ and $j_\alpha$. Note that we add one for the number of unknown variables because the constant $c$ is also an unknown variable. In the case of 3D space, we have a group of 48 elements $G_{48}$ generated by $c(i_\alpha, j_\alpha, k_\alpha)$ with a non-zero constant $c$ and three different natural numbers $i_\alpha$, $j_\alpha$ and $k_\alpha$, a group of 24 elements $G_{24}$ by $c(i_\alpha, i_\alpha, j_\alpha)$ or $c(i_\alpha, j_\alpha, 0)$, $G_{12}$ by $c(i_\alpha, i_\alpha, 0)$, $G_8$ by $c(i_\alpha, i_\alpha, i_\alpha)$, $G_6$ by $c(i_\alpha, 0, 0)$, or $G_1$ by $(0,0,0)$. Consequently,

$$n(\mathbf{v}_i) = \sum_{k \in I_D} kn_k \qquad (9)$$

where $n_i$ is the number of $G_i$ and $I_{D=3} = \{1,6,8,12,24,48\}$. The $n(\mathbf{v}_i)$ of the two-dimensional case can be also expressed by Formula (9) with $I_{D=2} = \{1,4,8\}$. The number of unknown variables is $1 + \sum_{k \in I_D} n_k$ so that we can construct a model by matching the numbers of unknown variables and equations such that

$$\Omega(m,D) = 1 + \sum_{k \in I_D} n_k . \qquad (10)$$

We should carefully choose the discrete velocities to satisfy the independency of the equations and to obtain solutions with real positive values of weight coefficient and speed. For example, models of fourth-order accuracy, *i.e.*, of the level of thermal compressible flow of Navier-Stokes, are explicitly given in Tables 2 and 3. In the cases of $D = 2$ and 3, the models having 8 and 10 groups are demanded because we have 9 and 11 equations by Formula (10), respectively, as in Table 1. The derived models have smaller sets of discrete velocities than the previously known models such as the 2D 37-velocity model and the 3D 107-velocity model [16] without losing accuracy. Of course, we can obtain the lower-order accuracy models including such as the 3D 15- and 19-velocities ones as well [24].

| No. | group | representative velocity | weight coefficient |
|---|---|---|---|
| 1 | $G_1$ | $(0,0)$ | .161987 |
| 2 | $G_4$ | $c(1,0)$ | .143204 |
| 3 | $G_4$ | $c(1,1)$ | .0338840 |
| 4 | $G_4$ | $c(2,0)$ | .00556112 |
| 5 | $G_4$ | $c(2,2)$ | $8.44799 \times 10^{-5}$ |
| 6 | $G_4$ | $c(3,0)$ | .00113254 |
| 7 | $G_8$ | $c(2,1)$ | .0128169 |
| 8 | $G_4$ | $c(4,4)$ | $3.45552 \times 10^{-6}$ |

**Table 2**. Details of the 2D 33-velocity model with $c = .819381$.

| No. | group | representative velocity | weight coefficient |
|---|---|---|---|
| 1 | $G_1$ | $(0,0,0)$ | .206847 |
| 2 | $G_6$ | $c(2,0,0)$ | .00442257 |
| 3 | $G_{12}$ | $c(2,2,0)$ | .0333341 |
| 4 | $G_8$ | $c(2,2,2)$ | .0128902 |
| 5 | $G_6$ | $c(3,0,0)$ | .0287920 |
| 6 | $G_{12}$ | $c(3,3,0)$ | .00264319 |
| 7 | $G_8$ | $c(3,3,3)$ | .000927908 |
| 8 | $G_{24}$ | $c(2,2,5)$ | .00106078 |
| 9 | $G_{12}$ | $c(4,4,0)$ | .000804376 |
| 10 | $G_6$ | $c(5,0,0)$ | .00274697 |

**Table 3**. Details of the 3D 95-velocity model with $c = .421803$.

To demonstrate the accuracy and stability of the models derived here, we show a shock tube simulation performed by the 3D 95-velocity model. The dimension of the nodes of the shock tube is $X \times Y \times Z = 11 \times 11 \times 800$. The two homogeneous initial conditions $C_L = \{\rho = p = 4, \theta = 1, \mathbf{u} = \mathbf{0}\}$ and $C_R = \{\rho = p = 1, \theta = 1, \mathbf{u} = \mathbf{0}\}$ are applied to the left ($1 \leq Z \leq 400$) and the right ($401 \leq Z \leq 800$) half spaces where $p$ is dimensionless pressure. The boundary conditions on the $XZ$- and $YZ$-planes are periodic and those on the left and the right $XY$-planes are $C_L$ and $C_R$, respectively. Figure 1 shows the density and the temperature profiles along the central longitudinal axis of the shock tube. The excellent agreements of the values between the analytical solutions and the simulation results are observed on the plateaus. Note that the result of the shock tube simulation with the model having the third-order momentum accuracy shows slight mismatches with respect to the analytical solution on the plateaus [14]. The difference of the steepness on the shock front comes from the difference of the viscosity. The analytical solution of the Riemann problem, which deals with the zero viscosity flow, is steeper than the result of the simulation dealing with the viscous flow with $\omega = 1.5$.

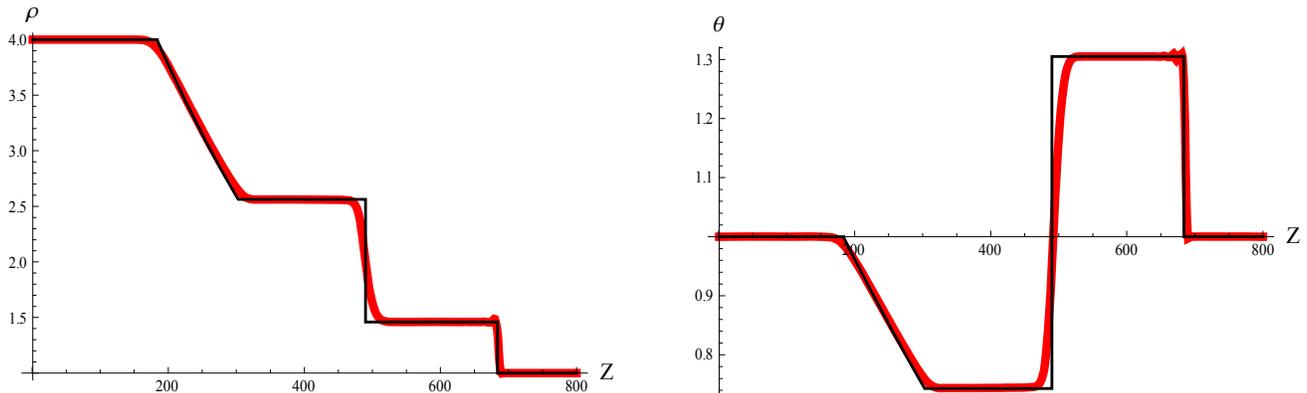

**FIGURE 1.** The density and the temperature profiles of the 3-dimensional shock tube simulation are shown on the left and the right, respectively. The analytical solution (thin black line) of the Riemann problem and the simulation results of $\omega = 1.5$ (thick red line) are shown.

We conclude this paper by noting that we can obtain the models with the levels of higher-order accuracy from the set of polynomial equations, Formula (5), expressed in a general form. The numbers of equations, discrete velocities, and unknown variables are given in Formulae (8), (9), and (10), for a given order of accuracy. As examples, explicit models of fourth-order accuracy are obtained and their numbers of discrete velocities are fewer in number than the previously known ones.

## ACKNOWLEDGMENTS

This work was partially supported by the KIST Institutional Program.

## REFERENCES

1. J. L. Schiff, Cellular Automata: a Discrete View of the World, Hoboken: Wiley, 2011
2. D. H. Rothman and S. Zaleski, Lattice-Gas Cellular Automata: Simple Models of Complex Hydrodynamics, Cambridge: Cambridge University Press, 1997, pp.5-11.
3. U. Frisch, B. Hasslacher, and Y. Pomeau, Phys. Rev. Lett. 56, 1505 (1986)
4. S. Wolfram, J. Stat. Phys. 45, 471 (1986)
5. J. W. Shim and R. Gatignol, Phys. Rev. E 81, 046703 (2010)
6. G. McNamara, G. Zanetti, Phys. Rev. Lett. 61, 2332 (1988)
7. F. Higuera, S. Succi, and R. Benzi, Europhys. Lett. 9, 663 (1989)
8. S. Chen, H. Chen, D. Martinez, and W. H. Matthaeus, Phys. Rev. Lett. 67, 3776 (1991)
9. H. Chen and W. H. Matthaeus, Phys. Rev. A 45, 5339 (1992)
10. Y. H. Qian, D. D'Humières, and P. Lallemand, Europhys. Lett. 17, 479 (1992)
11. X. Shan, X.-F. Yuan, and H. Chen, J. Fluid Mech. 550, 413 (2006)
12. W. P. Yudistiawan et al., Phys. Rev. E 82, 046701 (2010)

13. S. S. Chikatamarla and I. V. Karlin, Phys. Rev. Lett. 97, 190601 (2006)
14. J. W. Shim and R. Gatignol, Phys. Rev. E 83, 046710 (2011)
15. S. S. Chikatamarla and I. V. Karlin, Phys. Rev. E 79, 046701 (2009)
16. P. C. Philippi et al., Phys. Rev. E 73, 056702 (2006)
17. R. Surmas, C. E. Pico Ortiz, and P. C. Philippi, Eur. Phys. J. Special Topics 171, 81 (2009)
18. X. B. Nie, X. Shan, and H. Chen, EPL 81, 34005 (2008)
19. X. Shan, Phys. Rev. E 81, 036702 (2010)
20. R. Rubinstein and L.-S. Luo, Phys. Rev. E 77, 036709 (2008)
21. J. W. Shim, Phys. Rev. E 87, 013312 (2013)
22. J. W. Shim and R. Gatignol, Z. Angew. Math. Phys. (online first) DOI 10.1007/s00033-012-0265-1
23. G. E. Andrews, The Theory of Partitions, Cambridge: Cambridge University Press, 1998
24. Y. H. Qian, S. Succi, and S. A. Orszag, "Recent Advances in Lattice Boltzmann Computing" in Annual Reviews of Computational Physics III, edited by D. Stauffer, Singapore: World Scientific, 1995, pp. 195-242